\documentclass[%
 reprint,
 amsmath,amssymb,
 aps,
 showkeys
]{revtex4-2}

\usepackage{graphicx}
\usepackage{dcolumn}
\usepackage{bm}

\usepackage{xcolor}
\usepackage[version=4]{mhchem}

\begin{document}

\preprint{}

\title{On the distributed resistor-constant phase element transmission line in a reflective bounded domain}

\author{Anis Allagui$^*$}
\email{aallagui@sharjah.ac.ae}
\affiliation{Dept. of Sustainable and Renewable Energy Engineering, University of Sharjah, Sharjah, P.O. Box 27272, United Arab Emirates}
\altaffiliation[Also at ]{Center for Advanced Materials Research, Research Institute of Sciences and Engineering, University of Sharjah, Sharjah, P.O. Box 27272,  United Arab Emirates}
\affiliation{Dept. of Electrical and Computer Engineering, Florida International University, Miami, FL33174, United States}

%

\author{Enrique H. Balaguera}
\affiliation{
Escuela Superior de Ciencias Experimentales y Tecnología, Universidad Rey Juan
Carlos, C/ Tulip\'{a}n, s/n, 28933 M\'{o}stoles, Madrid, Spain
}

\author{Chunlei Wang}
\affiliation{Dept. of Mechanical and Aerospace Engineering, University of Miami, FL, United States}

\begin{abstract}

In this work we derive and study the analytical solution of the voltage and current diffusion equation for the case of a finite-length resistor-constant phase element (CPE) transmission line (TL) network that can represent a model for porous electrodes in the absence of any Faradic processes. The energy storage component is considered to be an elemental  CPE per unit length of impedance $z_c(s)={1}/{(c_{\alpha} s^{\alpha})}$ with constant parameters $(c_{\alpha},\alpha)$ instead of the ideal capacitor of impedance $z(s)={1}/{(c\, s)}$ usually assumed in TL modeling. The problem becomes a time-fractional diffusion equation for the voltage that we solve under galvanostatic charging, and derive from it a reduced impedance function of the form $z_{\alpha}(s_n)=s_n^{-\alpha/2}\coth({s_n^{\alpha/2}})$, where $s_n = j\omega_n$ is a normalized frequency.  We also derive the system's step response, and the distribution function of relaxation times associated with it. The analysis  can be viewed and used as a support  for the fractal finite-length Warburg model.

\end{abstract}

\maketitle


\section{Introduction}

The reduced impedance model given by  
$ z(s_n) ={  {s_n^{-1/2}}   } { \coth ( { s_n^{1/2}}   ) }$, 
  where $s_n = j \omega_n$ is a normalized angular frequency, is known as the reflective finite-length Warburg impedance model that represents the linear diffusion dynamics in  restricted thin electrodes \cite{bisquert2001theory, bisquert2000doubling, song2018electrochemical, moya2024low}. It has been introduced  by Franceschetti and MacDonald \cite{franceschetti1979diffusion} to describe the case of  diffusion of gas or metal species through a metallic electrode or along the electrode/electrolyte interface, and by Gabrielli et al. \cite{gabrielli1987impedance} to describe  reactions on polymer film coated electrodes. It can be shown that this  diffusion impedance transitions from that of the Warburg element at high frequencies, scaling as $s_n^{-1/2}$,  to that of a resistance in series with a capacitance at low frequencies  \cite{gabrielli1987impedance}. However,  experimental evidence  for many electrochemical systems shows clear deviations  from these characteristic features, including for instance the case where insertion reactions  are taking place\;\cite{farcy1990kinetic, gabrielli1987impedance, ho1980application, bisquert2000doubling, xu2021tunable, ohayon2024high}.   Cabanel et al.  \cite{cabanel1993determination} proposed a variant of the   restricted diffusion impedance function by introducing empirically  a dispersion coefficient $\alpha$ ($0<\alpha<1$) in the function so that it takes the form: $  {  {s_n^{-\alpha/2}}   } { \coth ( { s_n^{\alpha/2}}   ) }$. This model has   since been applied   to the study of a multitude of electrochemical systems, such as the
diffusion of hydrogen in a thin layer of \ce{H_xNb_2O_5} \cite{cabanel1993determination}, lithium-ion batteries \cite{profatilova2020impact}, supercapacitors \cite{ohayon2024high}, etc.

\begin{figure}[!t]
\begin{center}
\includegraphics[width=.4\textwidth]{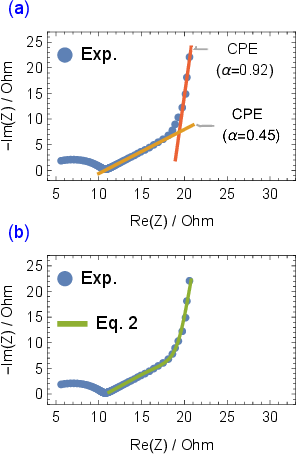}
\caption{Nyquist plot of experimental impedance data measured for a NEC/Tokin supercapacitor device on a Biologic VSP-300 electrochemical station from 1\,MHz down to 10\,mHz with $V_{ac}=\text{10\,mV rms}$. In solid lines we show fitting of the impedance data with the models given by Eqs.\;\ref{RsCPE} (the two straight lines in (a)) and\;\ref{cothx} (green line in (b))}
\label{fig0}
\end{center}
\end{figure}

\begin{figure*}[ht]
\begin{center}
\includegraphics[angle=-90,width=0.8\textwidth]{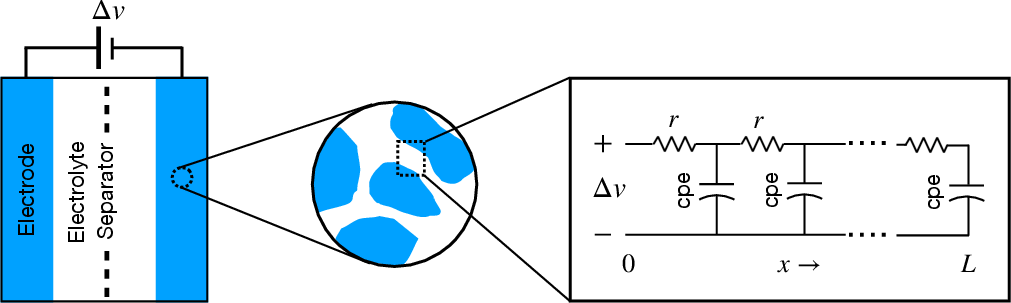}
\caption{Schematic of an electrified pore of a porous electrode in contact with an electrolyte modeled by a finite-length $R$-CPE transmission line model. }
\label{fig1}
\end{center}
\end{figure*}

As an example, we show in Fig.\;\ref{fig0}  the Nyquist impedance plot  of a commercial NEC/Tokin supercapacitor, model FG Series, rated 5.5\,V, 1.0\,F (part num. FG0H105ZF). The device was first maintained at a constant voltage of 2.7\,V for 5\,min to avoid undesired transient variations in its electrical response, and then subjected to a 10\,mV rms stepping sine excitation from the frequency of  1\,MHz down to 10\,mHz.  We recorded the impedance data at 10 frequency points per decade.  
Complex nonlinear least squares fitting of the very low-frequency branch from 10 to 20\,mHz (i.e. from the point of coordinates (20.59,\;22.09)\,Ohm to the point of coordinates (19.75,\;12.24)\,Ohm)   with a model consisting of a resistance in series with a constant phase element (CPE), i.e. of impedance:
\begin{equation}
Z_l(f)= R_s + \frac{1}{C_{\alpha}  (j 2 \pi f)^{\alpha}}
\label{RsCPE}
\end{equation} 
gives the line    in red color in  Fig.\;\ref{fig0}(a). This line is inclined with an angle of 83 deg. vs. the real axis, which  corresponds to an $\alpha$-value of  $\approx 0.92$. The values of the series resistance  $R_s$ and pseudocapacitance $C_{\alpha}$ are found to be $18.0$\,Ohm and 0.56\,F\,s$^{\alpha-1}$, respectively. 
Whereas the other inclined   line in the figure (in orange color) which is fitting the impedance data spanning the frequency range from 65\,mHz to 35\,Hz   is inclined with an angle of 40 deg., i.e. $\alpha \approx 0.45$. This is practically half of the values corresponding to the fit of the low-frequency segment of the data. Here the series resistance and pseudocapacitance are found to be $10.8$\,Ohm and 0.16\,F\,s$^{\alpha-1}$, respectively. 
Now fitting the low- and high-frequency impedance data together from 10\,mHz to 35\,Hz with the dimensional model corresponding to Eq.\;\ref{eqZan} below, i.e.:
\begin{equation}
Z_{\text{TL}}(f) = R_s + R_d \frac{ \coth (( j 2 \pi f \tau)^{\alpha/2}) } {( j 2 \pi f \tau)^{\alpha/2}}
\label{cothx}
\end{equation}
gives the parameters set: 
$R_s= 10.8$\,{Ohm}, 
$R_d = 24.2$\,Ohm,
$\tau=17.8$\,s, and
$\alpha=0.94$.   As one can appreciate from Fig.\;\ref{fig0}(b), the curved impedance data are nicely fitted with this model. We note that the value of the parameter $\alpha$ obtained from Eq.\;\ref{cothx} ($\alpha=0.94$)
 is practically equal to that obtained from fitting the low frequency branch of impedance using the $R_s$-CPE model of Eq.\;\ref{RsCPE} ($\alpha\approx 0.92$).

In this study, we demonstrate that this modified impedance function can be derived from the anomalous time-fractional diffusion equation  representing a distributed $R$-CPE transmission line (TL) in a bounded one-dimensional domain (Eq.\;\ref{eqDt} below).  
We recall that in a single pore of an electrified porous electrode filled with electrolyte, the electrolyte's resistance and the electrolyte-electrode interface double-layer capacitance can be viewed respectively as a distributed resistance and distributed capacitance over the the length of the pore \cite{de1963porous}. This equivalent ladder-like network model can be represented by cutting up the overall resistance and capacitance, and connecting the building blocks in a  TL   network (Fig.\;\ref{fig1}). In the limit of infinitely many infinitesimally small self-similar  circuit elements connected together,   the potential drop across the capacitors of the   TL network can be modeled by a diffusion-type   partial differential equation    \cite{pedersen2023equivalent}. That is:
\begin{equation}
\frac{\partial v(x,t)}{\partial t} = \frac{1}{rc} \frac{\partial^2 v(x,t)}{\partial x^2}, \quad 0<x< \infty,\;t \geqslant 0
\end{equation}
where $r$ and $c$ represent respectively  constant resistance/capacitance per unit length of the TL.

In a recent contribution by some of the co-authors \cite{R-CPE-TL}, this problem of diffusion has been generalized and studied for the case of a  CPE  acting as the capacitive energy storage element instead of an ideal capacitor. This is to account more generally  for the anomalous resistive-capacitive behavior observed in both frequency and time-domain  data on most porous electrodes used in practical applications (supercapacitors, fuel cells, batteries, etc.).   The current on the CPE is given by the fractional-order differential equation:
\begin{equation}
i_c(x,t) = C_{\alpha}\,
{}_0\text{D}_t^{\alpha} v_c(x,t)
\label{iCPE}
\end{equation}
 where  $C_{\alpha}$ is a pseudocapacitance in units of F\,s$^{\alpha-1}$, and    
 ${}_0\text{D}_t^{\alpha} $ is the Caputo differential  operator of order $\alpha$ ($0<\alpha \leqslant 1$) defined as:
\begin{equation}
{}_0\text{D}_t^{\alpha} f(t) := \frac{1}{\Gamma(m-\alpha)} \int_0^t (t-\tau)^{m-\alpha-1} f^{(m)}(\tau) d\tau
\end{equation}
Here $m\in \mathbb{N}$, $m-1< \alpha < m$ ($m=1$ in our case), $\Gamma(z) =\int_0^{\infty} u^{z-1} e^{-u} du,\,(\text{Re}(z)>0)$ is the gamma function, and $f^{(m)}(t)=d^m f(t)/dt^m$ is the $m^{th}$ derivative of $f(t)$ with respect to $t$. 
This leads in the frequency domain to the impedance of the CPE being \cite{lazanas2023electrochemical, allagui2023tikhonov, 10.1149/1945-7111/ac621e, allagui2021possibility, allagui2021inverse,cpe}:
\begin{equation}
Z_c(s) =  \frac{1}{C_{\alpha} s^{\alpha}}
\end{equation}
($s= j \omega$) which simplifies to that of an ideal capacitor when $\alpha=1$. The  problem of anomalous diffusion for the case of a semi-infinite $R$-CPE TL network becomes:
\begin{equation}
{}_0\text{D}_t^{\alpha} v(x,t) = \frac{1}{rc_{\alpha}} \frac{\partial^2 v(x,t)}{\partial x^2}, \quad 0<x<\infty,\;t \geqslant 0
\label{eqDt0}
\end{equation}
 where $r$ and $c_{\alpha}$ represent again a constant resistance/pseudo-capacitance per unit length.  
 The solution obtained for Eq.\;\ref{eqDt0} in response to a step voltage at $x=0$ and assuming zero current at infinity has been derived using the Laplace transform method, and provided in terms of the Fox's $H$-function \cite{R-CPE-TL}. We also derived the total impedance of the modified TL network, and  found that the system  itself behaves as a CPE of half the order of that of the constituting CPEs ($Z_{\text{TL}}(s)=  \sqrt{r/c_{\alpha} } s^{-\alpha/2}$). This can be viewed  as a generalization of the Warburg semi-infinite diffusion along $RC$-based TL, which is known to lead to an equivalent CPE of order $0.5$ \cite{huang2018diffusion}.
 
Here, we focus  on the solution of the same time-fractional diffusion along a self-similar $R$-CPE TL network, but  on a bounded one-dimensional domain $[0;L]$ instead (see Fig.\;\ref{fig1}).  The motivation is that in porous electrodes in contact with an electrolyte in electrochemical devices such as supercapacitors, batteries, capacitive desalination modules and electrochemical (bio)sensors \cite{janssen2021transmission, gupta2020charging, janssen2021locating, pedersen2023equivalent}, the generalized TL diffusion equation on a finite length interval is in practice more relevant to the study of their  responses  than the cases of infinite or semi-infinite length \cite{song2018electrochemical}. We make use of combined transforms using the finite Fourier cosine transform on the spatial variable and the Laplace transform on the time variable in the diffusion equation \cite{agrawal2002solution}. The voltage drop is found as an infinite series of the Mittag-Leffler function which can also be expressed in terms of the Fox's $H$-function.  We also derive the impedance   of the network and found it to be indeed in the desired  form of 
${ {   {s_n^{-\alpha/2}}   } \coth (  { s_n^{\alpha/2}}  ) }$. From this impedance we derive by inverse Laplace transform the system’s step response, and by another iteration of inverse Laplace transform the 
distribution function of relaxation times of the Debye type associated with it.

\section{Theory}

\subsection{Statement of the  problem}

We consider in this study the time-fractional partial differential equation (PDE) of 1D diffusion over the bounded domain $[0; L]$:
\begin{equation}
{}_0\text{D}_t^{\alpha} v(x,t) = \frac{1}{rc_{\alpha}} \frac{\partial^2 v(x,t)}{\partial x^2}, \quad 0<x<L,\;t > 0
\label{eqDt}
\end{equation} 
  Furthermore, we consider the following  boundary Neumann conditions \cite{posey1966theory}:
\begin{align}
-\frac{1}{r_0} \left. \frac{\partial v(x,t)}{\partial x}\right|_{x=0} &= i_0, \quad t > 0   \label{eqBC1} \\
\left. \frac{\partial v(x,t)}{\partial x}\right|_{x=L} &= 0, \quad t > 0   \label{eqBC2} 
\end{align}
where $i_0$ is a constant current, 
and the   initial condition: 
\begin{equation}
v(x,t=0) = 0, \quad  0< x < L
\label{eqIC}
\end{equation}
Again, this TL model describes the case of galvanostatic  charging of a uniform and homogeneous pore in a porous conductive material completely  filled with an electrolyte, where the capacitive behavior is represented by a CPE of constant parameters (Fig.\;\ref{fig1}).

\subsection{Solution by operational method} 

First we represent the voltage $v(x,t)$ as\;\cite{tomovski2013exact}:
\begin{equation}
v(x,t) = v_1(x,t) + v_2(x,t)
\end{equation}
The function $v_2(x,t)$ is chosen to satisfy the nonhomogeneous     boundary conditions given by Eqs.\;\ref{eqBC1} and\;\ref{eqBC2}, i.e.:
\begin{align}
  \left. \frac{\partial v_2(x,t)}{\partial x}\right|_{x=0} &= - r_0 i_0, \quad t > 0   \label{eqBC1v2} \\
\left. \frac{\partial v_2(x,t)}{\partial x}\right|_{x=L} &= 0, \quad t > 0   \label{eqBC2v2} 
\end{align}
which results in\;\cite{tomovski2013exact}:
\begin{equation}
v_2(x,t)= - r_0 i_0 x + \frac{ r_0 i_0 x^2 }{2 L }
\end{equation}
 The problem of $v_1(x,t)$ is
 \begin{equation}
{}_0\text{D}_t^{\alpha} v_1(x,t) = \frac{1}{rc_{\alpha}} \frac{\partial^2 v_1(x,t)}{\partial x^2}, \quad 0<x<L,\;t > 0
\end{equation}
with the  homogeneous boundary conditions
\begin{align}
  \left.\frac{\partial v_1(x,t)}{\partial x}\right|_{x=0} &=0, \quad t > 0   \label{eqBC1v3} \\
\left.\frac{\partial v_1(x,t)}{\partial x}\right|_{x=L} &= 0, \quad t > 0   \label{eqBC2v3} 
\end{align}
and the initial condition
\begin{equation}
v_1(x,t=0) = r_0 i_0 x - \frac{ r_0 i_0 x^2 }{2 L }, \quad  0< x < L
\label{eqICvq}
\end{equation}
 We now apply the finite Fourier  cosine transform  to the problem of $v_1(x,t)$ with respect to the variable $x$. 
 The finite Fourier  cosine transform is 
 defined for a function $f(x)$ in the interval $[0;L]$ as \cite{strandhagen1944use,  roettinger1947generalization, al2018finite}
\begin{equation}
    \bar{f}(k) = \mathcal{F}_c[{f}(x); k ] = \int\limits_0^{L} {f}(x) \cos(k \pi x / L) dx
\end{equation}
where $k=1,2,\ldots$. This is derived from the Fourier cosine series for $f(x)$:
\begin{equation}
f(x) =\frac{a_0}{2} + \sum\limits_{k=1}^{\infty} a_n \cos(k \pi x/L)
\end{equation} 
Its inverse  transform is given by: 
\begin{equation}
  f(x) = \mathcal{F}^{-1}_c[\bar{f}(k);x ]  = \frac{1}{L} \bar{f}(0) + \frac{2}{L} \sum\limits_{k=1}^{\infty} \bar{f}(k) \cos(k \pi x / L)  
\end{equation}
After   taking into account the boundary conditions (Eqs.\;\ref{eqBC1v3} and\;\ref{eqBC2v3}), the transformed equation becomes:
\begin{equation}
    {}_0\text{D}_t^{\alpha} \bar{v}_1(k,t) 
   + \frac{(k\pi)^2}{r c_{\alpha} L^2}  \bar{v}_1(k,t)
    =0
   \label{eqFST}
\end{equation}
We carry on by taking the Laplace transform  
of both sides of Eq.\;\ref{eqFST} with respect to $t$, which leads to the diffusion equation in the $(k,s)$-domain to be:
\begin{equation}
    \tilde{\bar{v}}_1(k,s) = \frac{ \bar{g}_1(k,0) s^{\alpha-1}}{s^{\alpha} + {(k\pi)^2}/{(r c_{\alpha} L^2)}}
\end{equation}
where:
\begin{align}
\bar{g}_1(k,0)&= \mathcal{F}_c[v_1(x,t=0) ; k ]  \\ 
&= r_0 i_0 L^2 \frac{-2  k \pi +(2 + k^2\pi^2) \sin(k\pi) }{2 k^3 \pi^3}
\end{align} 
We recall that the Laplace transform is defined for a function $f(t)\; (t>0)$ as
\begin{equation}
\tilde{f}(s)=\mathcal{L}\{f(t);s\} = \int\limits_0^{\infty} e^{-st} f(t) dt,\;\; \text{Re}(s)>0
\end{equation}
and the inverse Laplace transform is defined as:
\begin{equation}
    f(t) = \mathcal{L}^{-1}\{\tilde{f}(s);t\} = \frac{1}{2 \pi i} \int\limits_{\gamma - i \infty}^{\gamma + i \infty} e^{st} \tilde{f}(s) ds
\end{equation} 
Also, the Laplace transform  of the Caputo time-fractional derivative of order $\alpha$ is given by:
\begin{equation}
\mathcal{L}\left[{}_0\text{D}_t^{\alpha} f(t); s \right]= s^{\alpha} \tilde{f}(s) - \sum\limits_{k=0}^{m-1} s^{\alpha-k-1} f^{(k)}(0^+)
\end{equation} 

Now for the inversion steps, we start  first by applying the inverse Laplace transform using the Prabhakar formula \;\cite{prabhakar1971singular}:
\begin{equation}
    \mathcal{L}^{-1}\left\{ \frac{s^{\beta-1}}{s^{\alpha}+a} ; t \right\} = t^{\alpha-\beta} \text{E}_{\alpha, \alpha - \beta +1} \left( -a t^{\alpha} \right)
    \label{eqMLLT}
\end{equation}  
where  $\text{E}_{a,b}^{c} ( z )$ is the three-parameter Mittag-Leffler function defined by\;\cite{mathai2009h}:
 \begin{equation}
\text{E}_{a,b}^{c} ( z ) = \sum\limits_{k=0}^{\infty} \frac{(c)_k}{\Gamma(a k + b)} \frac{z^k}{k!} \quad (a,b, c \in \mathbb{C}, \mathrm{Re}({a})>0)
\label{eqML}
\end{equation}
 with  $(c)_k = c(c+1)\ldots(c+k-1) =\Gamma(c+k)/\Gamma(c)$.
 Thus,  we obtain:
\begin{align}
    \bar{v}_1(k,t) 
    &= \bar{g}_1(k,0) \text{E}_{\alpha} \left( - \frac{(k\pi)^2 t^{\alpha}}{r c_{\alpha} L^2}  \right) \label{barxtML2}
\end{align}  
Then, by inverse finite  Fourier cosine transform, 
  we obtain the voltage part $v_1(x,t)$ as the infinite series:
   \begin{align}
v_1(x,t) &= \frac{r_0 i_0 L}{3} 
\nonumber \\ & 
-  {2 r_0 i_0 L} \sum\limits_{k=1}^{\infty} \frac{\cos(k\pi x/L)}{(k\pi)^2} \text{E}_{\alpha} \left( - \frac{(k\pi)^2 t^{\alpha}}{r c_{\alpha} L^2}  \right)
 \end{align}
 Similar results were shown also by Luchko\;\cite{luchko2012initial, luchko2010some}   using the method of separation of variables. 
 The total voltage in the TL network is therefore given by:
  \begin{align}
v(x,t) &= \frac{r_0 i_0 L}{3} - r_0 i_0 x + \frac{ r_0 i_0 x^2}{2 L} + \frac{ r_0 i_0 }{r c_{\alpha} L} \frac{t^{\alpha}}{\Gamma(1+\alpha)} \nonumber \\
& - {2 r_0 i_0 L} \sum\limits_{k=1}^{\infty} \frac{\cos(k\pi x/L)}{(k\pi)^2} \text{E}_{\alpha} \left( - \frac{(k\pi)^2 t^{\alpha}}{r c_{\alpha} L^2}  \right) 
\label{eq:vxtH}
\end{align} 
with\;\cite{tomovski2013exact}:
\begin{equation}
\frac{ r_0 i_0 }{r c_{\alpha} L} \frac{t^{\alpha}}{\Gamma(1+\alpha)}
=
\frac{1}{2}\, {}_0\text{D}_t^{-\alpha} \left( \frac{2}{L} \int\limits_0^{L} \frac{r_0 i_0}{r c_{\alpha} L} dx \right) 
\end{equation}
in which the term in the integrand being
\begin{equation}
\frac{r_0 i_0}{r c_{\alpha} L}=    \frac{1}{rc_{\alpha}} \frac{\partial^2 v_2(x,t)}{\partial x^2} - {}_0\text{D}_t^{\alpha} v_2(x,t)
\end{equation}
 
 For $\alpha=1$, the voltage reduces to the ideal $RC$-based TL case (some terms are missing in \cite{posey1966theory}):
  \begin{align}
v(x,t) &= \frac{r_0 i_0 L}{3} - r_0 i_0 x + \frac{ r_0 i_0 x^2}{2 L} + \frac{r_0 i_0\,t}{r c L}  \nonumber \\
&  -{2\, r_0 i_0 L} \sum\limits_{k=1}^{\infty} \frac{\cos(k \pi x / L)}{(k \pi   )^2}   \exp\left( -\frac{(k \pi  )^2 t}{r c_1 L^2} \right)  
 \end{align} 
  We verify  using\;\cite{gradshteintable2015}:
\begin{equation}
\sum\limits_{n=1}^{\infty} \frac{\cos(n x)}{n^2} = \frac{\pi^2}{6} - \frac{\pi x}{2} + \frac{x^2}{4}, \quad 0 \leqslant x \leqslant 2\pi
\end{equation}
that the initial voltage at $t=0$ is indeed zero.

The   current $i(x,t)$ along the TL can be obtained from the gradient of the voltage  via:
\begin{equation}
i(x,t) = -\frac{1}{r} \frac{\partial v(x,t)}{\partial x}
\label{eqI1}
\end{equation}
which gives:
 \begin{align}
i(x,t) &= \frac{r_0 i_0 (L-x)}{r L} 
\nonumber \\ &
-\frac{2 r_0 i_0 }{r} 
\sum\limits_{k=1}^{\infty} \frac{\sin(k \pi x / L)}{(k \pi)}   \text{E}_{\alpha} \left( -\frac{(k \pi )^2 t}{r c_{\alpha} L^2} \right)
\label{eq:ixt}
 \end{align} 
We verify that for $x=0$ and $r=r_0$, we retrieve the input current $i_0$ (boundary condition given by Eq.\;\ref{eqBC1}), and for $x=L$,  the current is zero.  
 
\subsection{Derivation of impedance function}

The impedance of the electrode/electrolyte system is computed from the ratio of the Laplace transform of the voltage by that of the current at the surface $x=0$ \cite{R-CPE-TL,pedersen2023equivalent}, i.e.: 
 \begin{equation}
    Z_{\text{TL}}(s) = \frac{ \mathcal{L}\left[ v (x=0,t); s \right]}{i_0/s}
        \label{eqZ}
\end{equation}
With the use of the Laplace transform formula for an the Mittag-Leffler function (Eq.\;\ref{eqMLLT})  
and knowing that $\coth(z)$ can be expressed by the series representation:
 \begin{equation}
\coth(z) = \frac{1}{z} + 2 z  \sum\limits_{k=1}^{\infty} \frac{1}{z^2 + (k \pi)^2}   
\end{equation}
the definition given in Eq.\;\ref{eqZ} leads to the reduced impedance function:
 \begin{equation}
\frac{Z_{\text{TL}}(s)}{r_0 L }
= \frac{ \coth\left( \sqrt{ r c_{\alpha} L^2 s^{\alpha}}  \right) }{  \sqrt{ r c_{\alpha} L^2 s^{\alpha}}   }
\label{eqZTLs}
\end{equation}
This is clearly a direct generalization of the case of $RC$-based TL network having $\alpha=1$ \cite{bisquert2001theory, pedersen2023equivalent}. Taking the normalized angular frequency $s_n= s ( r c_{\alpha} L^2 )^{1/\alpha} $, we can rewrite the impedance function in the form:
 \begin{equation}
\frac{Z_{\text{TL}}(s_n)}{r_0 L }
=  {  {s_n^{-\alpha/2}}   } { \coth ( { s_n^{\alpha/2}}   ) }
\label{eqZan}
\end{equation}

Furthermore,  we can proceed by applying one round of   inverse Laplace transform using the Prabhakar formula (Eq.\;\ref{eqMLLT})
 or using
 \begin{equation}
\mathcal{L}^{-1} \left\{   \frac{ \mathcal{L} \{ v(x=0,t) ;s \} }{i_0/s}    ;t \right\} = \frac{1}{i_0}  \frac{\partial v(x=0,t)}{\partial t}
\end{equation}
while taking into account the zero initial condition (Eq.\;\ref{eqIC}) to  obtain
 \begin{align}
      \mathcal{L}^{-1}&\{ Z_{\text{TL}}(s)  ;t \}  
      \nonumber \\ &  
= \frac{r_0 t^{\alpha-1}}{r c_{\alpha} L \,\Gamma(\alpha)}   + \frac{2 r_0}{r c_{\alpha} L} \sum\limits_{k=1}^{\infty} t^{\alpha-1} \text{E}_{\alpha,\alpha} \left( -\frac{(k \pi  )^2 t^{\alpha}}{r c_{\alpha} L^2} \right) \label{eq45}
      \end{align}
  Note that  the $m^{th}$ ($m \in \mathbb{N}$) derivative  of $t^{\beta-1} E_{\alpha,\beta}^{\gamma} (a t^{\alpha})$ is given by \cite{giusti2020practical}:
\begin{equation}
\frac{d^m}{dt^m}  t^{\beta-1} E_{\alpha,\beta}^{\gamma} (a t^{\alpha}) = t^{\beta-1-m} E_{\alpha,\beta-m}^{\gamma} (a t^{\alpha})
\end{equation}
and
  \begin{equation}
\text{E}_{\alpha,0}(z) = z \text{E}_{\alpha,\alpha}(z)
\end{equation} 
 Eq.\;\ref{eq45} corresponds to the TL system's
response  function commonly obtained in experimental chronopotentiometric experiments.
We verify that for $\alpha=1$, we obtain:
\begin{equation}
\mathcal{L}^{-1}\{ Z_{\text{TL}}(s)  ;t \}=\frac{r_0}{r c_1 L} 
\vartheta_3\left[0,\exp\left({-\frac{ \pi^2 t}{r {c_1} L^2 }} \right)\right]\end{equation}
where $\vartheta_3(z,q)$ is the Jacobi elliptic theta function defined as \cite{et3}:
\begin{equation}
\vartheta_3(z,q) = 1+ 2 \sum\limits_{k=1}^{\infty} q^{k^2} \cos(2 k z) 
\end{equation} 

By applying another round of inverse Laplace transform  to Eq.\;\ref{eq45}, we can derive the distribution function of relaxation times $g(\tau)$ following the Debye $RC$ elemental process, as shown   in \cite{allagui2024procedure}.  We use   the  inverse Laplace transform of the Mittag-Leffler function  \cite{van2020mittag} (formula\;\#92):
\begin{equation}
x^{1-1/\alpha} \text{E}_{\alpha,\alpha}(-x) = \frac{\sin(\alpha \pi)}{\pi} \int\limits_0^{\infty} \frac{u^{\alpha} e^{-x^{1/\alpha} u} du}{1+2 u^{\alpha} \cos(\alpha \pi) + u^{2\alpha}}  
\end{equation}
where $x$ is replaced with $b s^{\alpha}$ and $u$ is replaced with $b^{-1/\alpha} v$, which gives:
\begin{align}
s^{\alpha-1} & \text{E}_{\alpha,\alpha}(-b s^{\alpha}) = 
\nonumber \\ & 
 \frac{\sin(\alpha \pi)}{\pi} \int\limits_0^{\infty} \frac{v^{\alpha} e^{-s v}  dv}{b^{1/\alpha+1}+2 b^{1/\alpha} v^{\alpha} \cos(\alpha \pi) +  b^{1/\alpha-1} v^{2\alpha}} 
\end{align}
Then, from Eq.\;\ref{eq45} the following expression for $g(\tau)$ is derived:
\begin{align}
g(\tau) &=  \frac{  \sin (\pi  \alpha ) \tau^{\alpha-1}}{\pi r {c_{\alpha}} L^2 }  
\nonumber \\
& \times \left( 1 +  \sum _{k=1}^{\infty } \frac{2\, }{b_k^{1/\alpha+1} \tau^{2\alpha} + 2 b_k^{1/\alpha} \tau^{\alpha} \cos(\alpha \pi) + b_k^{1/\alpha-1}   } \right)
\label{eq:gtau2}
\end{align}
 where $b_k=(k \pi)^2/(r c_{\alpha}L^2)$ and $0<\alpha<1$. 
This means that the  impedance function of a self-similar $R$-CPE TL network can be expressed as the integral
\begin{equation}
\frac{Z_{\text{TL}}(s)}{r_0 L} = \int\limits_{0}^{\infty} \frac{g(\tau)}{{1+ s \tau}} d\tau
\end{equation}
using the reduced impedance of an $RC$ model as a basis function (i.e. $(1+s\tau)^{-1}$), with $g(\tau)$ (Eq.\;\ref{eq:gtau2}) specifying  the distribution of the $RC$ time constants $\tau$. Note that $g(\tau)$ can also be derived using the Laplace transform property of the Fox's $H$-function\;\cite{allagui2024procedure}, knowing that the generalized Mittag-Leffler function   can be expressed in terms of the $H$-function as\;\cite{mathai2009h}:
\begin{equation}
\text{E}_{a,b}^{c} ( z )= 
\frac{1}{\Gamma(c)}
H^{1,1}_{1,2}\left[ -z  \left|
\begin{array}{l}
(1-c,1) \hfill   \\
(0,1), (1-b,a)    \\
\end{array}
\right.\right],\; \text{Re}(a)>0
\end{equation}
and the inverse Laplace transform of $H$-function is 
\begin{align}
\mathcal{L}^{-1} 
&\left[ 
  u^{-\rho} 
H_{p,q}^{m,n}\left[ a  u^{\sigma}\left|
\begin{array}{c}
(a_p,A_p)  \\
(b_q,B_q)  \\
\end{array}
\right.\right]; t
\right] \label{eq:ILT} \\ \nonumber
 &= t^{\rho-1} 
H_{p+1,q}^{m,n}\left[ a t^{-\sigma}\left|
\begin{array}{l}
(a_p,A_p), \ldots, (a_1,A_1), (\rho,\sigma)  \\
(b_1,B_1),\ldots,(b_q,B_q) \hfill \\
\end{array}
\right.\right]
\end{align}
Following our procedure in\;\cite{allagui2024procedure}, we can show that:
\begin{equation}
g(\tau)=\frac{  \tau^{\alpha-1} }{ r {c_{\alpha}} L^2 } \left( \frac{\sin (\pi  \alpha )}{\pi} + 2\sum_{k=1}^{\infty} 
H_{p,q}^{m,n}\left[ b_k  \tau^{\alpha} \left|
\begin{array}{l}
(0,1), (1-\alpha,\alpha)  \\
(0,1), (1-\alpha,\alpha)  \\
\end{array}
\right.\right]
\right) 
\label{eq:gtauH}
\end{equation}


\section{Results and discussion}

\begin{figure*}[!t]
\begin{center}
\includegraphics[width=0.9\textwidth]{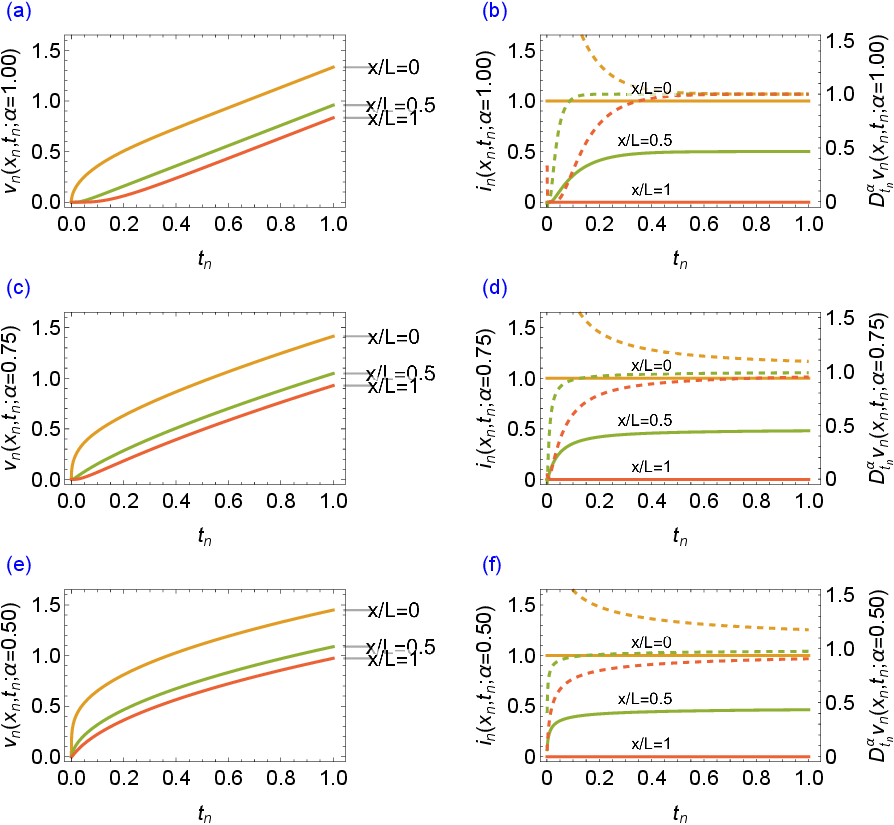}
\caption{Time-domain voltage and current response  of an $R$-CPE TL network in a bounded domain $[0;\,L]$ under galvanostatic charging:  (a) Plots of voltage $v_n(x_n,t_n)$ given by Eq.\;\ref{eqvn}, and  (b) plots of the corresponding current $i_n(x_n,t_n)$ given by Eq.\;\ref{eq:inxt} (solid lines) and the current through the CPE part given by Eq.\;\ref{eq:caputoDvnxt} (dashed lines) for $c_{\alpha}=1$  vs. $t_n=t/(r c_{\alpha} L^2)^{1/\alpha}$
for $x/L=0,\,0.5,\,1.0$ and $\alpha=1$. The same is repeated in (c) and (d) respectively for $\alpha=0.75$, and in (e) and (f) for $\alpha=0.5$}
\label{fig3}
\end{center}
\end{figure*}

Several quantities have been obtained from the analysis  of the bounded anomalous diffusion problem with the Caputo time-fractional derivative given by:
\begin{align}  
\begin{cases}
{}_0\text{D}_t^{\alpha} v(x,t) = \cfrac{1}{rc_{\alpha}} \cfrac{\partial^2 v(x,t)}{\partial x^2}, \quad & 0<x<L,\;t > 0 \\
 \left. \frac{\partial v(x,t)}{\partial x}\right|_{x=0} = -r_0 i_0, \quad & t > 0   \\
\left. \cfrac{\partial v(x,t)}{\partial x}\right|_{x=L} = 0, \quad & t > 0  \\
v(x,t=0) = 0, \quad & 0< x < L
\end{cases}
\end{align}  
This includes the voltage and current as a function of time and position,  as well as the input impedance at $x=0$.  In this section we study numerically these electrical characteristics   for different values of the dispersion coefficient $\alpha$ of the constituting CPE.   
  
With the dimensionless variables $t_n=t/(r c_{\alpha} L^2)^{1/\alpha}$ \cite{R-CPE-TL} and $x_n=x/L \in [0;1]$, the voltage  given by Eq.\;\ref{eq:vxtH} is rewritten normalized to the voltage $ r_0 i_0 L$ as:      
\begin{align}
v_n(x_n ,t_n)
&= \frac{1}{3} - x_n + \frac{x_n^2}{2} + \frac{t_n^{\alpha}}{\Gamma(1+\alpha)} \nonumber \\
&   - 2 \sum\limits_{k=1}^{\infty}  \frac{{\cos(k \pi x_n)}}{(k\pi)^2}  
     \text{E}_{\alpha} \left( -  {(k\pi)^2 t_n^{\alpha}}  \right)
    \label{eqvn}
  \end{align} 
 The current $i(x,t)$ (Eq.\;\ref{eq:ixt})  normalized with respect to $i_0(r_0/r)$, and denoted $i_n(x,t)$, takes the form:
 \begin{align}
i_n(x,t) =  (1-x_n) 
- 2 
\sum\limits_{k=1}^{\infty} \frac{\sin(k \pi x_n)}{(k \pi)}  \text{E}_{\alpha} \left( -  {(k\pi)^2 t_n^{\alpha}}  \right)
\label{eq:inxt}
 \end{align} 
 and the Caputo time derivative of order $\alpha$ of $v_n(x_n ,t_n)$ is given by:
\begin{equation}
{}_0\text{D}_{t_n}^{\alpha} v_n(x_n,t_n)
  = 1
   + 2 \sum\limits_{k=1}^{\infty}  {\cos(k \pi x_n)}  
      \text{E}_{\alpha} \left( -  {(k\pi)^2 t_n^{\alpha}}  \right)
      \label{eq:caputoDvnxt}
  \end{equation}  
  The latter is proportional to  the current on the CPE parts of the TL network (see Eq.\;\ref{iCPE}) at different times and locations.
  
  Numerical simulations of Eq.\;\ref{eqvn} vs. $t_n$ for 
  the positions $x_n=x/L=0,\,0.5,\,1.0$ on the TL network and for $\alpha=1.0,\,0.75,\,0.5$ are given in the first column of Fig.\;\ref{fig3}. 
  The infinite  sum is truncated to its first 10 terms. Note that for the particular case of $\alpha=0.5$, the Mittag-Leffler function reduces to $\text{E}_{1/2}(z) = e^{z^2} \text{erfc}(-z)$ where $\text{erfc}(z)$ is the complementary error function, and again for $\alpha=1$, we have $\text{E}_1(z) = e^{z}$.  We also have to mention that having the same scale of the $x$-axis for all plots in  Fig.\;\ref{fig3} implies that we are assuming   $t/(r c_{\alpha} L^2)^{1/\alpha} $ is the same  for any value of $\alpha$ (i.e.  $(c_{\alpha})^{1/\alpha}=\text{cst}$ for $\alpha=1.0,\,0.75,\,0.5$). Therefore caution should be exercised when drawing direct conclusions on  the effect of $\alpha$ on the results.

For the ideal case of $\alpha=1$, Posey and Morozumi\;\cite{posey1966theory} reported overall similar results to those we have in Fig.\;\ref{fig3}(a), but not exactly the same  as some terms are missing in their original solution to the problem. We also observe a delay of how the voltage picks up vs. time for increasing values of $x/L$. Away from $t_n=0$, the voltage slope is quasi-linear for all values of $x/L$ along the TL network. For less-than-one values of $\alpha$ (Figs.\;\ref{fig3}(c) and\;\ref{fig3}(e)), the voltage vs. time follow power-law like profiles.

%
%


\begin{figure}[!t]
\begin{center}
\includegraphics[width=0.45\textwidth]{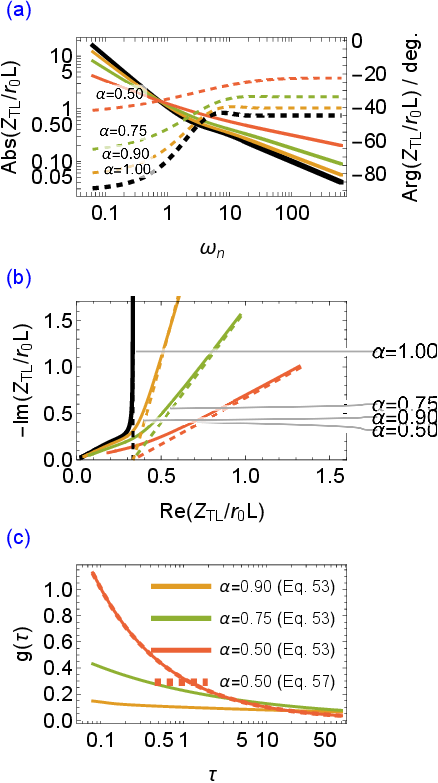}
\caption{Plot of the impedance function given by Eq.\;\ref{eqZan} for different values of $\alpha$ in terms of (a) magnitude vs. frequency (solid lines), and phase vs. frequency (dashed lines), and (b) imaginary vs. real part of impedance (dashed lines represent $ {1}/{3} +  {s_n^{-\alpha}}$ at the limit of $s_n \to 0$). In (c) we show plots of the associated distribution function of relaxation  times $g(\tau)$  (Eq.\;\ref{eq:gtau2} for $\alpha=0.90,0.75,0.50$ and Eq.\;\ref{eq:gtauH} for $\alpha=0.50$) vs $\tau$}
\label{fig4}
\end{center}
\end{figure}

Plots of  the respective currents $i_n(x,t)$ (Eq.\;\ref{eq:inxt})
 are shown in the second column of Fig.\;\ref{fig3} for the same values of $\alpha$ (solid lines). In any case, for $x/L=0$ we have always  the current   equal to $i_0$ as stated by the  condition on this boundary (Eq.\;\ref{eqBC1}), and the same is for the zero current at $x/L=1$ .  
 On the second $y$-axis of the   figures, we show plots of the Caputo time derivative of order $\alpha$ of $v_n(x_n ,t_n)$ (i.e. Eq.\;\ref{eq:caputoDvnxt}, dashed lines),   
which indicates saturation of the current on the CPE part of the TL network as time increases. The convergence of the current towards an asymptotic is faster as the  value of $\alpha$ is increased from 0.5 to 0.75 to 1.0. 

Finally, we recall again that the main objective of this work is to derive the impedance function given by Eq.\;\ref{eqZan} (i.e. $ Z_{\text{TL}}(s_n) \propto   {  {s_n^{-\alpha/2}}   } { \coth ( { s_n^{\alpha/2}}   ) }$), which was done by considering the diffusive transport of voltage and current  along an $R$-CPE-based TL network and not an $RC$-based one. This gives at least a physical motivation to  
Cabanel et al.'s   \cite{cabanel1993determination}  generalization of the   standard restricted diffusion impedance function (i.e. $  {  {s_n^{-1/2}}   } { \coth ( { s_n^{1/2}}   ) }$)  in which they  introducing empirically  the coefficient $\alpha$ without mathematical justification. 
 In Fig.\;\ref{fig4} we present simulation plots of  Eq.\;\ref{eqZan}   for  values of $\alpha$ between 1.0 and 0.5. The frequency range covered for the normalized angular frequency $\omega_n$ ($s_n= i \omega_n$) is $\pi/10$ to $200\,\pi$. The impedance plots are shown in terms of magnitude vs. frequency (Fig.\;\ref{fig4}(a)), phase vs. frequency (Fig.\;\ref{fig4}(b)), and (negative) imaginary vs. real part of the function (Fig.\;\ref{fig4}(c)). As discussed above, the impedance function shows clearly two separate regimes when looking at the Nyquist complex plots  in Fig.\;\ref{fig4}(c). For $s_n \to 0$,  the leading order of Eq.\;\ref{eqZan}  is  
$ {1}/{3} +  {s_n^{-\alpha}}$, which is marked in the figure by dashed straight lines for the different values of $\alpha$.  These lines form angles of $\alpha\pi/2$ with the real axis, and all intersect at the real axis at $\text{Re}(Z_{\text{TL}}/r_0 L)=1/3$.    For the ideal   case, we have a vertical line as it should be \cite{janssen2021locating}, which indicates pure capacitive behavior. Whereas at high frequencies, the impedance tends to $s_n^{-\alpha/2}$, which  corresponds to lines inclined from the real axis with half the inclination angle observed for the low frequency data ($\alpha \pi/4$). 
For $\alpha=1$, we obtain the classical case of 45 deg. inclined Warburg element for semi-infinite linear diffusion. 

We conclude by showing  in Fig.\;\ref{fig4}(c)  simulation plots  of the  derived distribution functions of relaxation time $g(\tau)$ (Eq.\;\ref{eq:gtau2}, up to the first 10 terms of the summation)  vs. $\tau$, for the $\alpha$-values of 0.90, 0.75  and 0.50. The time parameter $(r c_{\alpha}L^2)$ is set to one. 
It is clear that the distribution of relaxation times, as it should be, is narrower as $\alpha$ is increased toward one, and vice versa  it gets wider as $\alpha$ is decreased. 
 We also show for illustration 
$g(\tau)$ given by Eq.\;\ref{eq:gtauH} for $\alpha=0.5$, also up to the first 10 terms of the summation. The two curves  are in excellent  agreement with each other.


\section{Conclusion}
 
In this work, we showed that the modified impedance model ${ {   {s_n^{-\alpha/2}}   } \coth (  { s_n^{\alpha/2}}  ) }$ obtained empirically by inserting  the dispersion coefficient $\alpha$  into the classical theory ${ {   {s_n^{-1/2}}   } \coth (  { s_n^{1/2}}  ) }$ that describes bounded diffusion, can be derived exactly from the (Caputo) time-fractional diffusion equation (Eq.\;\ref{eqDt}). This corresponds to a homogeneous TL model consisting of distributed $R$s and CPEs of constant parameters. Importantly, we also derived by inverse Laplace transform the system's response to a step function (Eq.\;\ref{eq45}), and by another round of inverse Laplace transform the distribution function of relaxation times of the Debye type (Eq.\;\ref{eq:gtau2} or\;\ref{eq:gtauH}). These formulas are provided in terms of   the Mittag-Leffler function, which is a special function frequently encountered in fractional calculus. The results of this study can find many practical applications for  the fundamental  analysis of   electrochemical   systems in the time and frequency domains, as well as other mass or heat problems obeying Eq.\;\ref{eqDt}.

\section*{Acknowledgements}

C.W. acknowledges the funding support by the National Science Foundation,  awards \#2423124 and \#2412500.

\section*{References}


%

\end{document}